# Relativistic heavy-ion physics

*G. Herrera Corral**
CERN, Geneva, Switzerland

**Abstract**
The study of relativistic heavy-ion collisions is an important part of the LHC research programme at CERN. This emerging field of research focuses on the study of matter under extreme conditions of temperature, density, and pressure. Here we present an introduction to the general aspects of relativistic heavy-ion physics. Afterwards we give an overview of the accelerator facility at CERN and then a quick look at the ALICE project as a dedicated experiment for heavy-ion collisions.

## 1 Introduction

The study of relativistic heavy-ion collisions started in the 1970s at the Bevalac, Lawrence Berkeley National Laboratory, where a transport line was built to bring heavy ions from Hilac (Heavy ion linear accelerator) to the Bevatron. The Bevatron at LBNL is best known for the antiproton, discovered there in the 1955 by O. Chamberlain and E. Segré. The so-called Bevalac accelerated nuclei at about 1 *A* GeV/*c*. The demonstration that excited nuclear matter could be studied gave birth to research programmes at Brookhaven National Laboratory and at the European Organization for Nuclear Research (CERN). The Alternating Gradient Synchrotron (AGS) at Brookhaven National Laboratory in the United States accelerated silicon ions up to 15 *A* GeV. In Europe the Super Proton Synchrotron (SPS, CERN) produced a 60 *A* GeV beam of oxygen and then increased the energy to 200 *A* GeV.

Nowadays, research is conducted at the Relativistic Heavy Ion Collider (RHIC). This accelerator was completed in 1999 at Brookhaven National Laboratory in the United States. RHIC collides nuclear beams at 100 *A* GeV, i.e., at ten times more energy than at the SPS. At RHIC, four experiments are taking and analysing data: BRAHMS, PHENIX, PHOBOS, and STAR.

High-energy heavy-ion collisions involve large amounts of energy. RHIC accelerates gold nuclei at 100 GeV/nucleon, which means that each nucleus carries energy

$$100\ \text{GeV} \times 197\ \ \text{nucleons} = 19.7\ \ \text{TeV}.$$

In the centre of mass, these interacting ions deliver 39.4 TeV. The Large Hadron Collider at CERN will reach

$$\sqrt{s} = 1200 \quad \text{TeV}$$

in lead–lead interactions.

In high-energy collisions of protons and/or electrons, the energy available in the beam goes into a point interaction. In heavy-ion interactions, however, an enormous amount of energy is deposited in a small region of space and in a very short time. In this region the density of energy is so large that it may favour the appearance of new forms of matter. The search for these new forms of matter is the central objective of heavy-ion physics.

The energy density of nuclei with atomic number $A$ in normal conditions is given by

$$\varepsilon = \frac{A \times nucleon_{mass}}{V_{nuclear}}, \qquad \text{where} \qquad V_{nuclear} = \frac{4}{3}\pi (r_0 A^{1/3})^3.$$

* On sabbatical leave from Physics Department, CINVESTAV, Mexico City, Mexico.

A typical value of energy density for nuclear matter is $\varepsilon = 0.14$ GeV/fm$^3$. The energy densities reached at relativistic heavy-ion collisions are above 1 GeV /fm$^3$, i. e., 10 times larger than normal nuclei densities.

The future of these studies is now moving to CERN where the ALICE experiment is being prepared to study relativistic heavy-ion collisions at the highest energy ever. As mentioned above, the LHC will provide beams of lead at energies 30 times greater than at RHIC. The CMS and ATLAS experiments at the LHC will also study heavy-ion interactions in addition to their rich programme on proton–proton collisions.

Here we shall give an introduction to the new and exciting field of relativistic heavy-ion collisions. We take a quick historical look at Hagedorn's first predictions. We quickly go through Glauber's model to understand the way phenomena are experimentally evaluated and measured. We then explain the concept of energy density.

In order to introduce the QCD phase space diagram, we shall study the MIT bag model. This model provides an easy way to grasp general ideas before a more formal approach can be taken. With these tools we can discuss some of the probes and signatures that will uncover the appearance of a quark–gluon plasma. Finally we shall comment on the Large Hadron Collider as well as on the ALICE experiment which is dedicated to the study of ion–ion collisions at CERN.

## 2 Hagedorn limiting temperature

Rolf Hagedorn was the first to point out the possibility of a transition of ordinary matter into a plasma of quarks and gluons. He developed statistical physics methods and applied them to particle production in high-energy collisions. He observed that the measured density of hadron states grows exponentially, i.e.,

$$\frac{d\rho}{dm} \approx m^a e^{m/m_0} \tag{1}$$

where $m$ represents the mass of the observed hadrons and $a$ is a parameter [1]. In 1965 Hagedorn showed that this exponential behaviour implies a limiting temperature which he understood as a melting point of hadrons. Indeed, the number of states with energy in an interval between $E$ and $E + dE$ can be written [2] as

$$dn(E) \approx dE \int_0^E pE dm \frac{d\rho}{dm} e^{-E/kT} .$$

Introducing here the expression given in Eq. (1), and using $p^2 = E^2 - m^2$, one obtains

$$dn(E) \approx dE \int_0^E m^a e^{m/m_0} e^{-E/kT} \sqrt{E^2 - m^2}\, E dm .$$

Henceforth,

$$= E^{a+3} dE \int_0^1 z^a e^{zE/m_0} e^{-E/kT} \sqrt{1 - z^2}\, dz$$

with $m = zE$. Substituting $z = \cos(\varphi)$

$$= E^{a+3} e^{-E/kT} dE \int_0^{\pi/2} \cos^a(\varphi) \sin^2(\varphi) e^{E\cos(\varphi)/m_0} d\varphi \tag{2}$$

assuming $E / m_0 >> 1$, we may approximate Eq. (2) by

$$\approx E^{a+3} e^{-E/kT} dE \int_0^{\pi} \cos^a(\varphi) \sin^2(\varphi) e^{E\cos(\varphi)/m_0} d\varphi.$$

So that the integral can be calculated,

$$= E^{a+3} e^{-E/kT} dE \sqrt{\pi} \frac{2m_0}{E} \frac{\sqrt{\pi}}{2} \frac{e^{E/m_0}}{\sqrt{2\pi E / m_0}}$$

and then simplified to

$$= E^{a+3} dE \sqrt{\frac{\pi m_0^3}{2E^3}} e^{(E/m_0 - E/kT)} \quad .$$

Henceforth, the total energy density $\int_0^{\infty} E dn(E)$ diverges for $kT > kT_0 = m_0$.

The conclusion is therefore that no higher temperatures are possible or some new physics must become relevant.

Figure 1 shows the mass spectrum $\rho_{\exp}(m) = \sum \nu_i \delta(m - m_i)$ with $\nu_i = (2J_i + 1)(2I_i + 1)2^{\lambda_i}$ where $J$ is the spin, $I$ the isospin, and $\lambda = 1$, when particles are different from antiparticles and $\lambda = 0$ when particles are identical to their antiparticles. Figure 1 is then a comparison of the logarithmic smoothed mass spectrum for the hadronic particles known today and previously. One can see that the new hadron resonances improve the exponential behaviour predicted by Hagedorn.

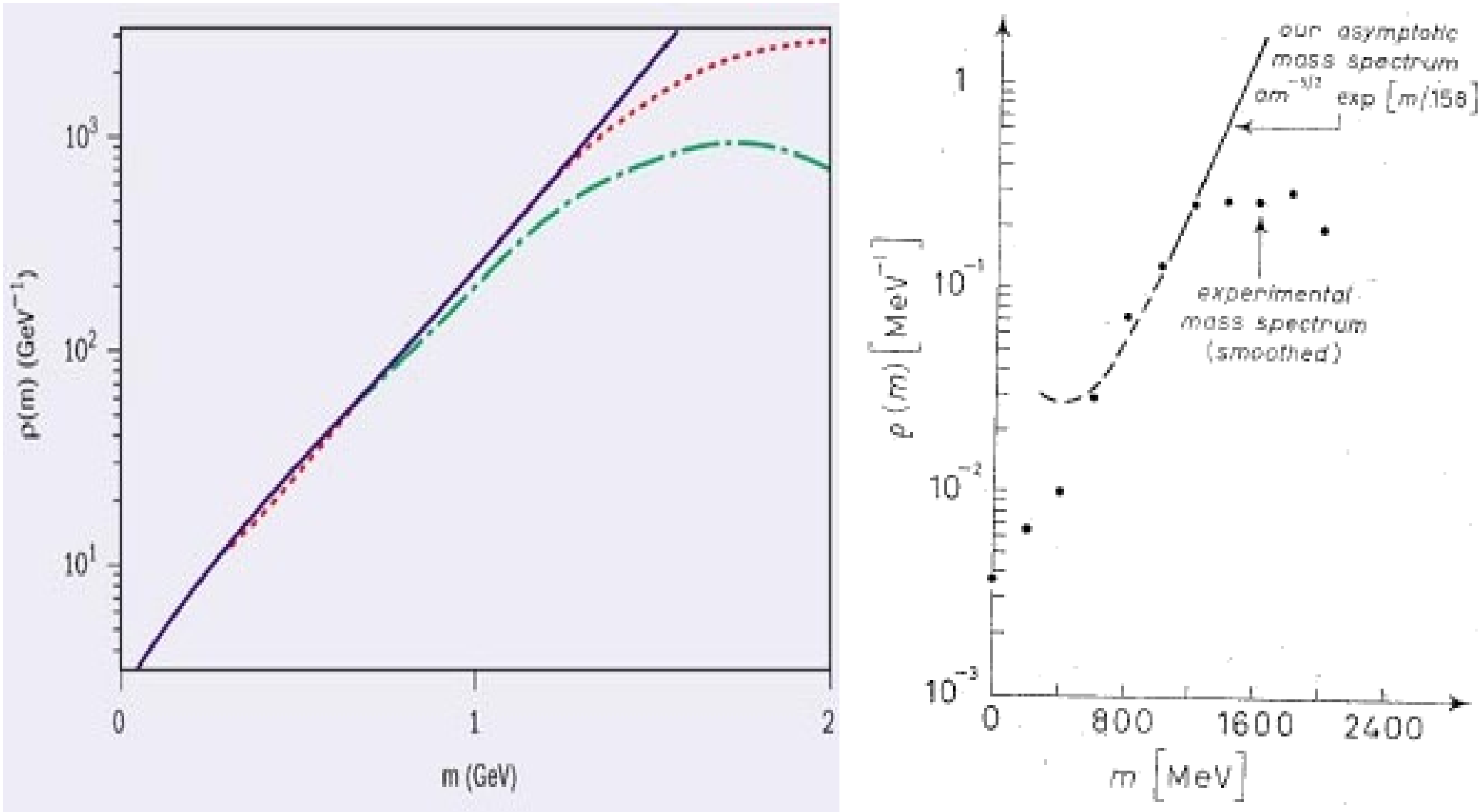

**Fig. 1:** On the left, a picture extracted from Ref. [3]; the solid blue line is the exponential fit to the smoothed hadron mass spectrum with present day information (represented here by the short-dashed red line). The long dashed green line corresponds to the Hagedorn spectrum obtained in 1967. On the right a similar picture extracted from the paper of Hagedorn from 1965 (see bibliography).

## 3 The Glauber model

The Glauber model [4], describes the interaction of two nuclei in terms of the interaction of the constituent nucleons. The model assumes the movement of the nucleus in a straight line and pictures the collision between the nuclei with a given impact parameter. In that sense it is a classical model of the interaction. It is widely used in heavy-ion collisions to describe interaction processes.

Figure 2 shows the geometry of a collision between nucleus $B$ and nucleus $A$. The probability of finding a baryon in the volume element $d\vec{b}_B dz_B$ of nucleus $B$ is $\rho(\vec{b}_B, z_B) d\vec{b}_B dz_B$.

A similar expression for a nucleus $A$ can be written. With this in mind, the probability element for having a baryon–baryon interaction when ions $A$ and $B$ collide with an impact parameter $\vec{b}$ is

$$dP = \underbrace{\rho_A(\vec{b}_A, z_A) d\vec{b}_A dz_A}\,\underbrace{\rho_B(\vec{b}_B, z_B) d\vec{b}_B dz_B}\,\underbrace{t(\vec{b} - \vec{b}_A - \vec{b}_B)\sigma_{in}}$$

*probability for finding a baryon in A* *in B. Probability for an inelastic collision.*

We define the thickness functions $T_A(\vec{b}_A) = \int dz_A \rho_A(\vec{b}_A, z_A)$ for nucleus $A$ and correspondingly $T_B(\vec{b}_B) = \int dz_B \rho_B(\vec{b}_B, z_B)$.

So we can write

$$T(\vec{b}) = \int d\vec{b}_A d\vec{b}_B T_A(\vec{b}_A) T_B(\vec{b}_B) t(\vec{b} - \vec{b}_A - \vec{b}_B). \quad (3)$$

With this, we can now write the probability for the occurrence of $n$ inelastic interactions when two nuclei $A$ and $B$ collide with an impact parameter $\vec{b}$ :

$$P(n,b) = \binom{AB}{n} [T(b)\sigma_{in}]^n [1 - T(b)\sigma_{in}]^{AB-n} .$$

The total probability of having an inelastic event in the collision of $A$ and $B$ is therefore

$$\frac{d\sigma^{AB}_{inel}}{db} = \sum_{n=1}^{AB} P(n,b) = 1 - [1 - T(b)\sigma_{in}]^{AB} . \quad (4)$$

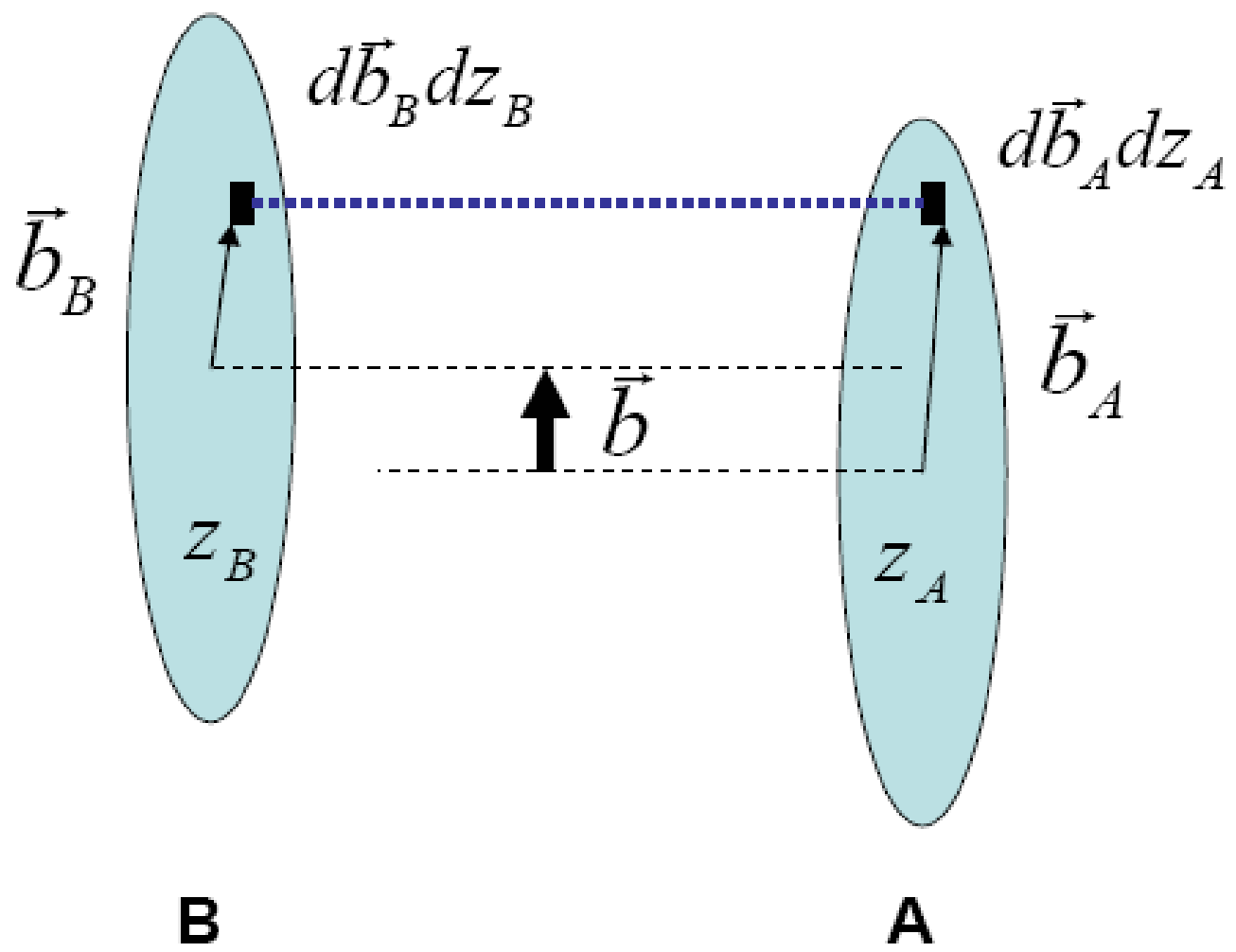

**Fig. 2:** Collision of nucleus $A$ with nucleus $B$ at impact parameter $\vec{b}$

From Eq. (4) one can see that the total inelastic cross section $\sigma^{AB}_{in}$ is

$$\sigma^{AB}_{inel} = \int db \{1 - [1 - T(b)\sigma_{in}]^{AB}\}. \quad (5)$$

One may approximate the thickness function $t$ in Eq. (3) with a Gaussian. For nuclei with small atomic number the density function can also be approximated by a Gaussian so that $T(b)$ in Eq. 3 can be written

$$T(b) = \exp(-b^2 / 2\beta^2) / 2\pi\beta^2 \cdot$$

The total inelastic cross-section is then

$$\sigma_{in}^{AB} = -2\pi\beta^2 \sum_{n=1}^{AB} \binom{AB}{n} \left( -\frac{\sigma_{in}^n}{n(2\pi\beta^2)^n} \right).$$

The simplest case of a proton–proton collisions with *n=1* is fulfilled in this approximation.

## 4 Energy density

The larger the number of nucleon–nucleon inelastic collisions the larger the energy deposited in the volume where those collisions take place.

Figure 3 shows two colliding ions *A´* and *B´*. The overlap area in the transverse direction is denoted with *A*. The volume formed by this area and a thickness length $\Delta z$ is then $A\Delta z$ . The number density of particles produced in that volume at $z = 0$ and at the time at which a quark–gluon plasma may form is given by

$$\frac{\Delta N}{A\Delta z} = \frac{1}{A}\frac{dN}{dy}\frac{dy}{dz}\bigg|_{y=0} = \frac{1}{A}\frac{dN}{dy}\frac{1}{\tau_0 \cosh y}\bigg|_{y=0} . \qquad (6)$$

Since $z = \tau \sinh y$ with $\tau = \sqrt{t^2 - z^2}$ , and where τ is the proper time. We evaluate $dy / dz$ in Eq. (6). This relation connects energy density and rapidity density. It was derived by Bjorken [5].

Considering that $E = m_T \cosh y$, with $E$ being the average energy per produced particle, the energy density at the moment of the collision is

$$\varepsilon_0 = \frac{\Delta N}{A\Delta z} m_T \cosh y. \qquad (7)$$

In this context, produced particles means everything appearing at rapidities intermediate between those of the original incoming nuclei.

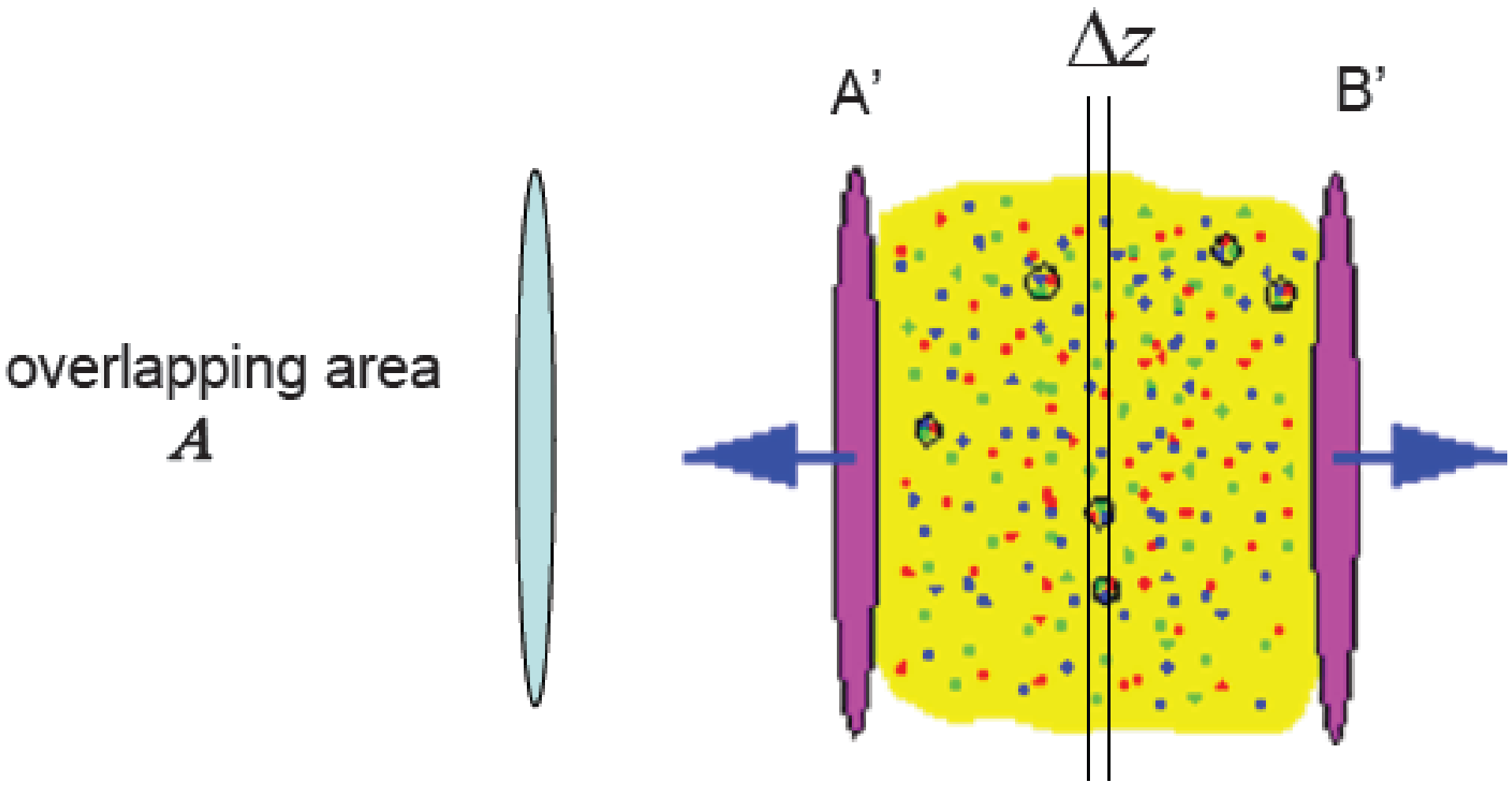

**Fig. 3:** Two colliding ions $A'$ and $B'$

Using Eq. (6) in Eq. (7) we obtain

$$\varepsilon_0 = \frac{m_T}{\tau_0 A} \left.\frac{dN}{dy}\right|_{y=0},$$

where $\tau_0$ is unknown. Bjorken estimated $\tau_0 = 1 fm/c$, however, the determination of the time scale at which the QGP is formed requires a knowledge of the dynamics behind.

## 5 The quantum chromodynamics phase diagram

In the MIT bag model [6], hadrons are thought of as closed containers of massless quarks which can be described by the Dirac equation. In the space time representation

$$(i\gamma^\mu p_\mu - m)\varphi = 0.$$

With $m = 0$, the equation becomes

$$\gamma p \varphi = 0,$$

i. e. ,

$$(\gamma^0 p^0 - \vec{\gamma}\cdot\vec{p})\varphi = 0.$$

In the Dirac representation of the $\gamma$ matrices,

$$\gamma^0 = \begin{pmatrix} I & 0 \\ 0 & -I \end{pmatrix} \qquad \gamma = \begin{pmatrix} 0 & \sigma \\ -\sigma & 0 \end{pmatrix}$$

the equation above can be written as

$$\begin{pmatrix} p^0 & -\vec{\sigma}\cdot\vec{p} \\ \vec{\sigma}\cdot\vec{p} & -p^0 \end{pmatrix}\begin{pmatrix} \varphi_+ \\ \varphi_- \end{pmatrix} = 0\cdot$$

These coupled equations $p^0\varphi_+ - \vec{\sigma}\cdot\vec{p}\varphi_- = 0$ and $\vec{\sigma}\cdot\vec{p}\varphi_+ - p_0\varphi_- = 0$ can be solved analytically. The lowest energy solution is

$$\varphi_+(\vec{r},t) = Ne^{-ip^0t} j_0(p^0r)\chi_+ \qquad \varphi_-(\vec{r},t) = Ne^{-ip^0t}\vec{\sigma}\cdot\hat{r}j_1(p^0r)\chi_- ,$$

in terms of the spherical Bessel functions $j_0$ and $j_1$ . Confinement can now be imposed by requiring the current flux through the spherical bag surface to be zero. This means that the normal component of the current $J_\mu = \varphi\gamma_\mu\varphi$ is equal to zero, i.e., $n^\mu\varphi\gamma_\mu\varphi = 0$, therefore $\bar{\varphi}\varphi = 0$.

This confinement condition means that

$$\bar{\varphi}\varphi\big|_{r=R} = \left[j_0(p^0R)\right]^2 - \hat{\sigma}\cdot\hat{r}\hat{\sigma}\cdot\hat{r}\left[j_1(p^0R)\right]^2 = 0.$$

That condition can be fulfilled if $p^0R = 2.04$ , which means that the energy of the quarks in the bag will be $E = 2.04N/R.$ For a bag under an external pressure $B$ , the energy of the quarks inside becomes

$$E = \frac{2.04N}{R} + \frac{4\pi}{3}R^3B.$$

The bag will be in equilibrium when

$$\frac{\partial E}{\partial R} = 0 , \qquad \text{i.e.,} \qquad 4\pi R^2B - \frac{2.04N}{R^2} = 0 .$$

Henceforth, a proton with three quarks (*N=3*) and radius *r* = 0.8 fm, will have external pressure $B^{1/4} = 1044 \times 197.3$, ($\hbar c = 197$ MeV fm), i.e. ,

$$B^{1/4} = 206 \text{ MeV.}$$

Let us now see what happens with a gas of quarks (fermions) and gluons (bosons) in thermal equilibrium. The total pressure of an ideal gas of quarks and gluons would be given by

$$P = \left[g_g + \frac{7}{8}(g_q + g_{\bar{q}})\right]\frac{\pi^2}{90}T^4 \qquad (8)$$

where $g_g = 8\times 2$ is the gluon degeneracy determined by the number of gluons and the two possible states. For the quarks we shall have $g_q = g_{\bar{q}} = N_{colors}\times N_{spin}\times N_{flavor}$ . The pressure can then be written

$$P = 37\frac{\pi^2}{90}T^4 .$$

When the pressure equals the bag pressure, i.e., $P = B$, the equation would give us the critical temperature at which the bag would break:

$$T_c = \left( \frac{90}{37\pi^2} \right)^{1/4} B^{1/4}. \qquad (9)$$

Figure 4 shows $T_c$ as obtained from Eq. (9) with $B^{1/4} = 206$ MeV.

In a similar way one could estimate a critical density and see that deconfinement may happen even at temperature $T = 0$. The number of quarks in a volume $V$ with momentum $p$ in the interval $dp$ is

$$N_q = \frac{g_q V}{(2\pi)^3} \int_0^{\mu_q} 4\pi p^2 dp = \frac{g_q V}{6\pi^2} \mu_q^3$$

so that the number density is

$$n_q = \frac{N_q}{V} = \frac{g_q}{6\pi^2} \mu_q^3.$$

The energy of these quarks in a volume $V$ is

$$E_q = \frac{g_q V}{(2\pi)^3} \int_0^{\mu_q} 4\pi p^3 dp = \frac{V g_q}{8\pi^2} \mu_q^4.$$

From the relation between pressure and energy $P_q = \frac{1}{3} \frac{E}{V}$, we obtain

$$P_q = \frac{g_q}{24\pi^2} \mu_q^4.$$

A change of state will occur when the pressure equals the bag pressure, i.e., $P_q = B$, this corresponds to a critical quark number density

$$n_q = 4 \left( \frac{g_q}{24\pi^2} \right)^{1/4} B^{3/4},$$

and taking baryons as groups of three quarks, the critical baryon number is therefore

$$N_q = \frac{4}{3} \left( \frac{g_q}{24\pi^2} \right)^{1/4} B^{3/4}.$$

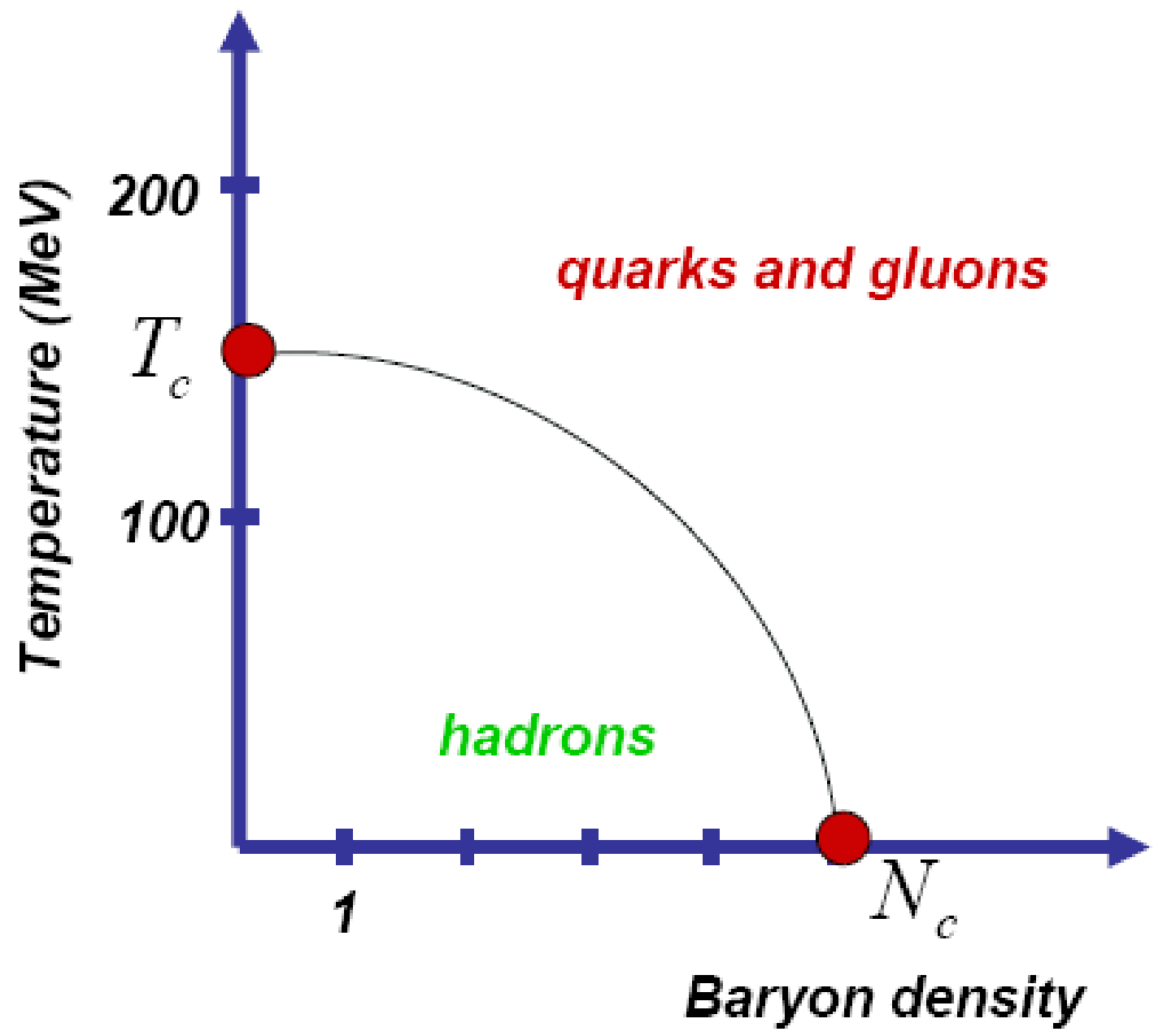

**Fig. 4:** QCD phase diagram. For high temperature, as discussed in the text, there is a critical temperature $T_c$. Beyond this temperature, the bag would break releasing quarks and gluons. Similarly a baryon density beyond $N_c$ would produce a phase transition.

We now take ordinary nuclear matter composed of *u* and *d* quarks only so that $g_q$ = $= 3_{colors} \times 2_{spin} \times 2_{flavor}$ and a bag pressure $B^{1/4} = 206$ MeV, the critical baryon number density at temperature $T = 0$, is

$$N_c = 0.72 / \text{fm}^3,$$

which corresponds to 5 times (see Fig 4) the normal nuclear density ($\varepsilon = 0.14$ GeV/fm$^3$, estimated in the introduction to this article.

## 6 Quark–gluon plasma probes and signatures

In order to know if a plasma of quarks and gluons has been created in the collision of ultrarelativistic heavy ions, we need observables. There are a number of ideas on what to look at to disentangle the short existence of a new state of matter. For lack of space, we shall not review all the probes and signatures considered by experiments nowadays, we shall only comment on some of them. The interested reader can then expand his knowledge from the bibliography recommended at the end of this article.

***Bose–Einstein correlations***

Two-particle correlations are among the most promising observables of the heavy-ion reaction to reveal the spacetime evolution (Fig. 5).

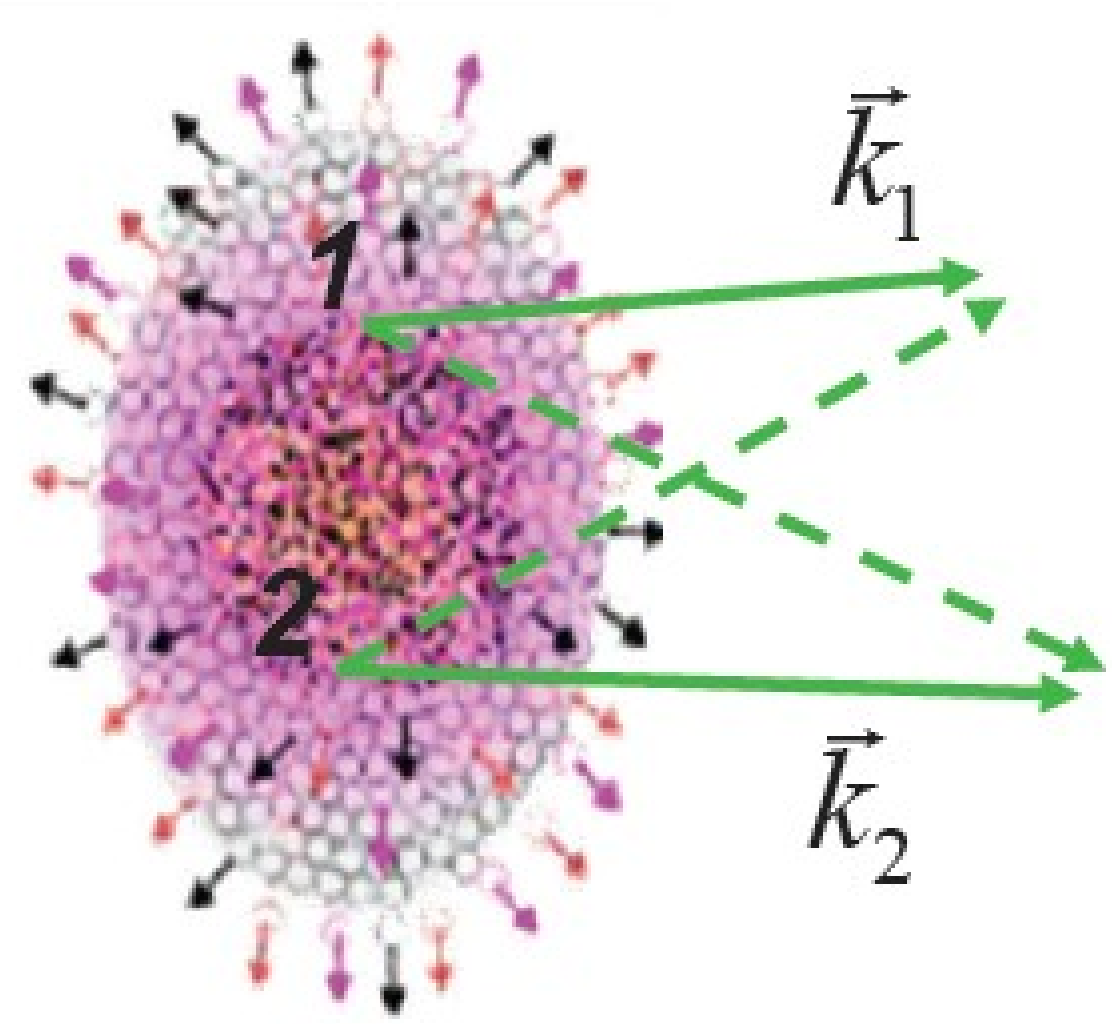

**Fig. 5:** Two identical particles are produced at different spacetime points 1 and 2, with momentum $k_1$ and $k_2$. Identical bosons obey Bose–Einstein statistics, so that quantum correlations are present and modify the phase space of the produced particles.

The wave function of the two particles produced at points $\vec{r}_1$ and $\vec{r}_2$ with momentum $\vec{k}_1$ and $\vec{k}_2$ can be written as

$$A_{\pi\pi} = e^{i\vec{k}_1\vec{r}_1 + i\vec{k}_2\vec{r}_2} + e^{i\vec{k}_1\vec{r}_2 + \vec{k}_2\vec{r}_1} .$$

The amplitude for the process is then

$$\left|A_{\pi\pi}\right|^2 = \left|e^{i\vec{k}_1\vec{r}_1 + i\vec{k}_2\vec{r}_2} + e^{i\vec{k}_1\vec{r}_2 + \vec{k}_2\vec{r}_1}\right|^2$$

$$\left|A_{\pi\pi}\right|^2 = 1 + 1 + e^{i(\vec{k}_1 - \vec{k}_2)\vec{r}_1} + e^{i(\vec{k}_2 - \vec{k}_1)\vec{r}_2}$$

$$\left|A_{\pi\pi}\right|^2 = 2 + 2\cos(qr)$$

where $q = \left|\vec{k}_1 - \vec{k}_2\right|$ and $r = \left|\vec{r}_1 - \vec{r}_2\right|$. Henceforth the probability of having two identical bosons, say pions, produced at two points in spacetime $\vec{r}_1$, $\vec{r}_2$ and with two momenta, $\vec{k}_1$, $\vec{k}_2$ divided by the probability of producing bosons independently, is given by

$$\frac{|A_{\pi\pi}|^2}{|A_\pi||A_\pi|} = 1 + \cos(qr)$$ .

We may introduce a probability density $\rho(x)$ for the pions to be produced at different points in spacetime. This amplitude would modify as follows:

$$A_{\pi\pi} = \rho(x)e^{i\vec{k}_1\vec{r}_1 + i\vec{k}_2\vec{r}_2} + \rho(x)e^{i\vec{k}_1\vec{r}_2 + \vec{k}_2\vec{r}_1}$$

with

$$\int \rho(x)dx = 1$$ .

In this case the ratio of amplitudes would contain the Fourier transform of the probability density $\rho(x)$, i.e.,

$$\frac{|A_{\pi\pi}|^2}{|A_\pi||A_\pi|} = 1 + |F(\rho)|^2 .$$

This ratio of amplitudes is the so-called two-particle correlation function.

$$C(k_1, k_2) = \frac{P(k_1, k_2)}{P(k_1)P(k_2)} = 1 + |F(\rho)|^2 .$$

Taking for example

$$\rho(x, y, z) = \frac{N}{4\pi^2 R_x R_y R_z \sigma_t} e^{-\left(\frac{x^2}{2R_x^2} + \frac{y^2}{2R_y^2} + \frac{z^2}{2R_z^2} + \frac{t^2}{2\sigma_t^2}\right)} ,$$

the correlation function would be

$$C(k_1 k_2) = 1 + F[\rho](q) = 1 + N' \exp\left[-\frac{1}{2}\left(R_x^2 q_x^2 + R_y^2 q_y^2 + R_z^2 q_z^2 + \sigma_t^2 q_t^2\right)\right].$$

By studying the two-particle correlation function one can measure the geometry of the particle production system.

Along these lines, one can use more sophisticated parametrizations. Figure 6 shows a common parametrization for the heavy-ion particle production environment and the results obtained by the PHENIX Collaboration using this particular geometry [7].

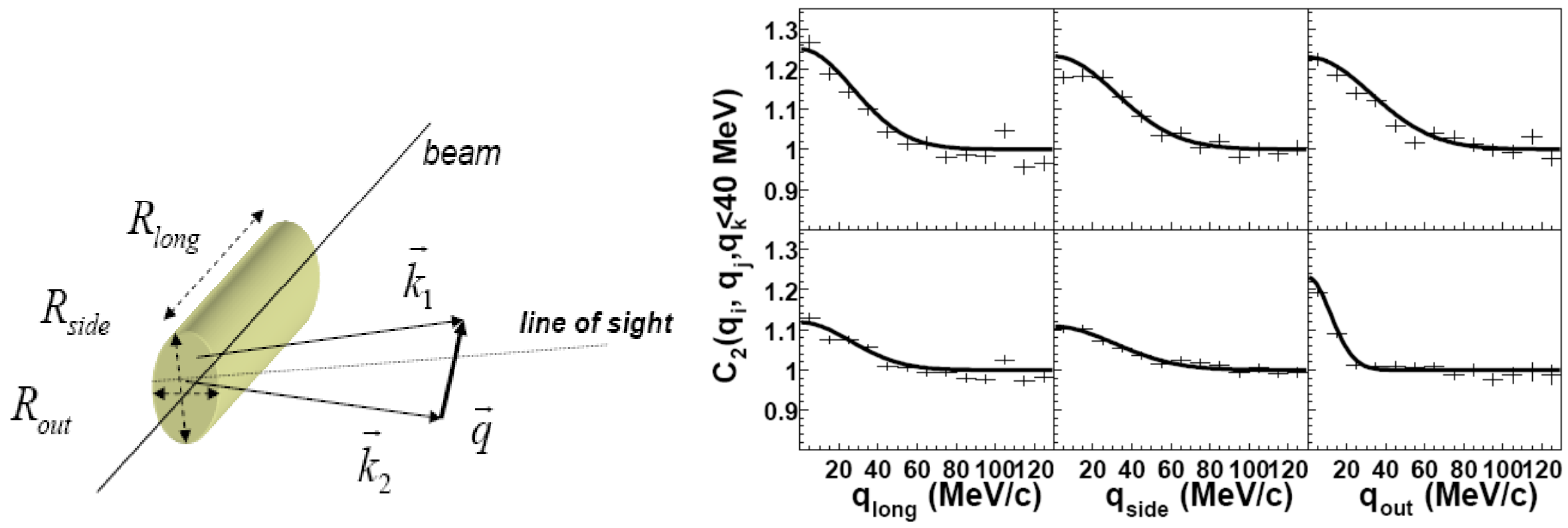

**Fig. 6:** On the left, a parametrization in terms of $R_{long}$ along the beam direction, $R_{out}$ along the *line of sight,* and $R_{side}$ perpendicular to the *line of sight*. The momenta of the pions are $\vec{k}_1$ and $\vec{k}_2$. On the right the experimental results by the PHENIX Collaboration [7] in terms of these parameters for pion pairs in the laboratory (top) and the pair centre-of-mass frame (bottom). The data is plotted as a function of one variable keeping the other two below 40 MeV/*c*.

On the other hand, if a quark–gluon plasma is produced, it will hadronize, populating the central rapidity region.

Considering $S_{QGP}$, the entropy of the quark–gluon plasma, and $S_{had}$, the entropy of the hadronization phase, then by the second law of thermodynamics

$$volume_{QGP} \times S_{QGP} \quad \leq \quad volume_{had} \times S_{had}.$$

Since $S_{had} < S_{QGP}$ then $V_{had} > V_{QGP}$.

Measuring the volume of the hadronization region by means of the two-particle correlation function one may say something about the production of a new state of matter. This is just an example of the ideas that have been considered in the frame of data coming from RHIC experiments. The source size extracted by fitting the correlation function to data grows with the event multiplicity and decreases with transverse momentum. However, the size and time of emission are anomalously large with respect to what has been suggested as signals for quark–gluon plasma formation. A better understanding of models and data is necessary.

***J/Ψ suppression***

The suppression of *J/Ψ* meson production was proposed in 1986 as a signature of a quark–gluon plasma [8]. It should be the manifestation of colour screening that would hinder *c* and anti-*c* quarks from binding to form a *J/Ψ* meson.

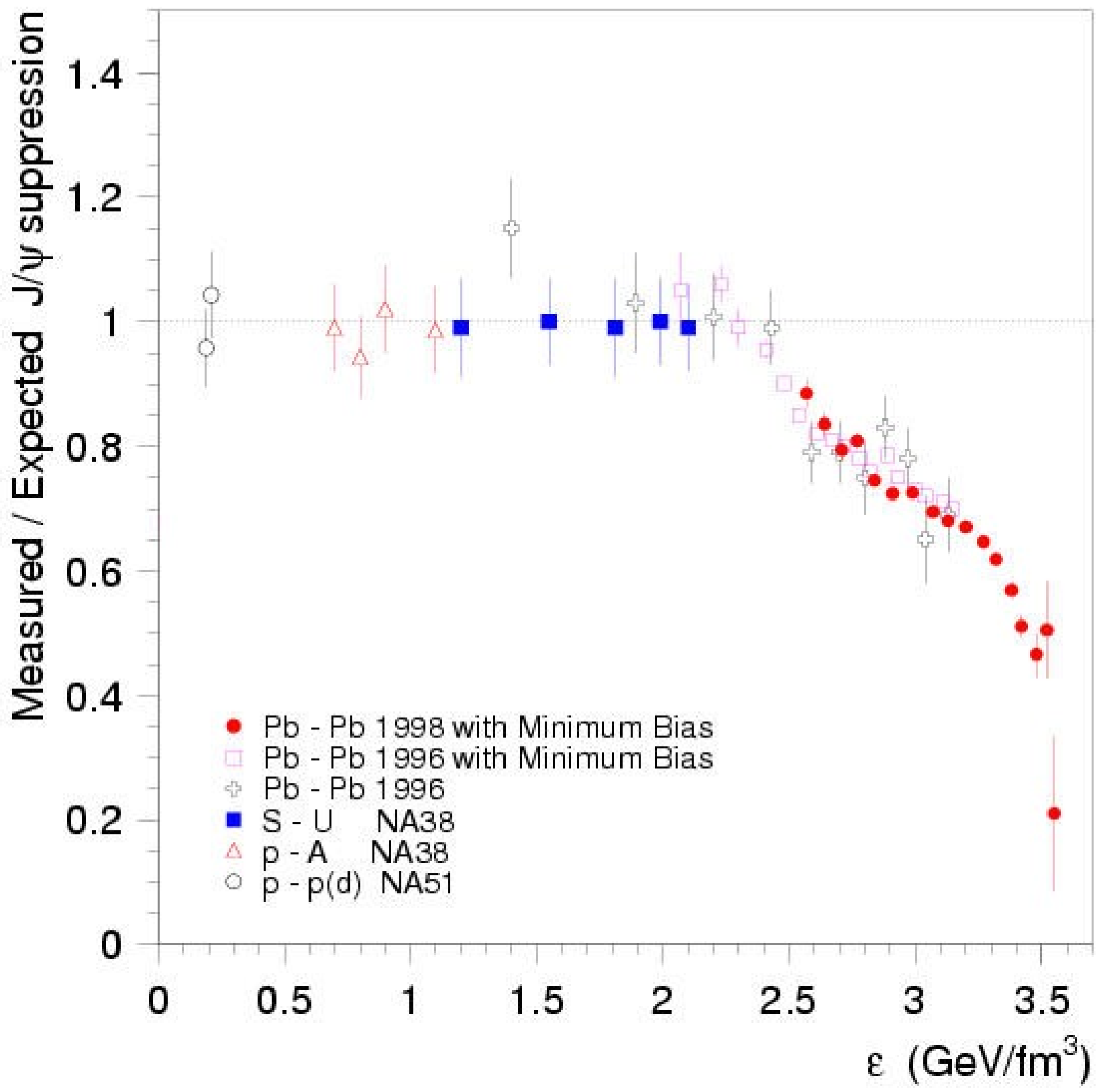

**Fig. 7:** Ratio of $J/\psi$ charmed mesons produced and expected, in several reactions and as a function of the energy density obtained in the reaction. Figure extracted from Ref. [9].

The first observation in experiment NA50 [9] at the SPS was explained as the result of inelastic interactions of these mesons with dense hadronic matter created in the collision. However, an anomalous suppression was observed later on by the NA50 and NA60 experiments.

The suppression has been a subject of study since then. A number of explanations like multiple scattering, gluon distribution changes, excited-state decays, heavy quark/gluon energy loss etc. have been provided. Further and more careful studies are needed.

Figure 7 shows the production of $J/\Psi$ mesons measured by several experiments in various reactions and as a function of the energy density reached in the collision. The measured cross-section for $J/\Psi$ production were divided by the values expected from nuclear absorption. One can see that in lead–lead interactions the production is suppressed according to the expected nuclear absorption for energy densities below $2.2\,\mathrm{GeV/fm}^3$. As higher energy densities are obtained, the suppression starts to become important. This may be the result of charmonium melting, i.e., a manifestation of QGP appearance.

***Jet quenching***

The phenomenon of jet quenching was proposed in 1982 by J. D. Bjorken [10] as the result of energy loss of quarks propagating through a quark–gluon plasma. In his paper [10] Bjorken says:

*High energy quarks and gluons propagating through a quark gluon plasma suffer differential energy loss via elastic scattering from quanta in the plasma. An interesting signature may be events in which*

*the hard collision occurs near the edge of the overlap region, with one jet escaping without absorption and the other fully absorbed.*

First evidence of parton energy loss has been observed at RHIC [11]. Observation of high $p_T$ hadron spectra and jet production in central Au–Au collisions and d–Au collisions confirmed the prediction of jet quenching.

Figure 8 show the azimuthal dependence of jets of particles. One sees clearly the presence of two jets in opposite directions in *proton–proton* and d–Au interactions. In Au–Au central collisions, however, one of the jets disappears. To obtain the plot in Fig. 8, one takes the highest transverse momentum track, which is between 4 and 6 GeV and then plots the tracks with transverse momentum in the interval $2\text{ GeV} < p_T < p_T^{trigger}$ associated with the azimuth $\Delta\phi$.

The strong suppression of pion production at $p_T$ up to 20 GeV has been observed at PHENIX [12] while direct photons which do not carry colour charge are not suppressed. The pions are generated by a fragmenting quark which does interact with the surrounding via its colour charge.

The magnitude of the measured suppression at high $p_T$ and jet-like angular correlations in central Au–Au collisions suggest that the initial energy density of the created medium is significantly larger than normal nuclear density.

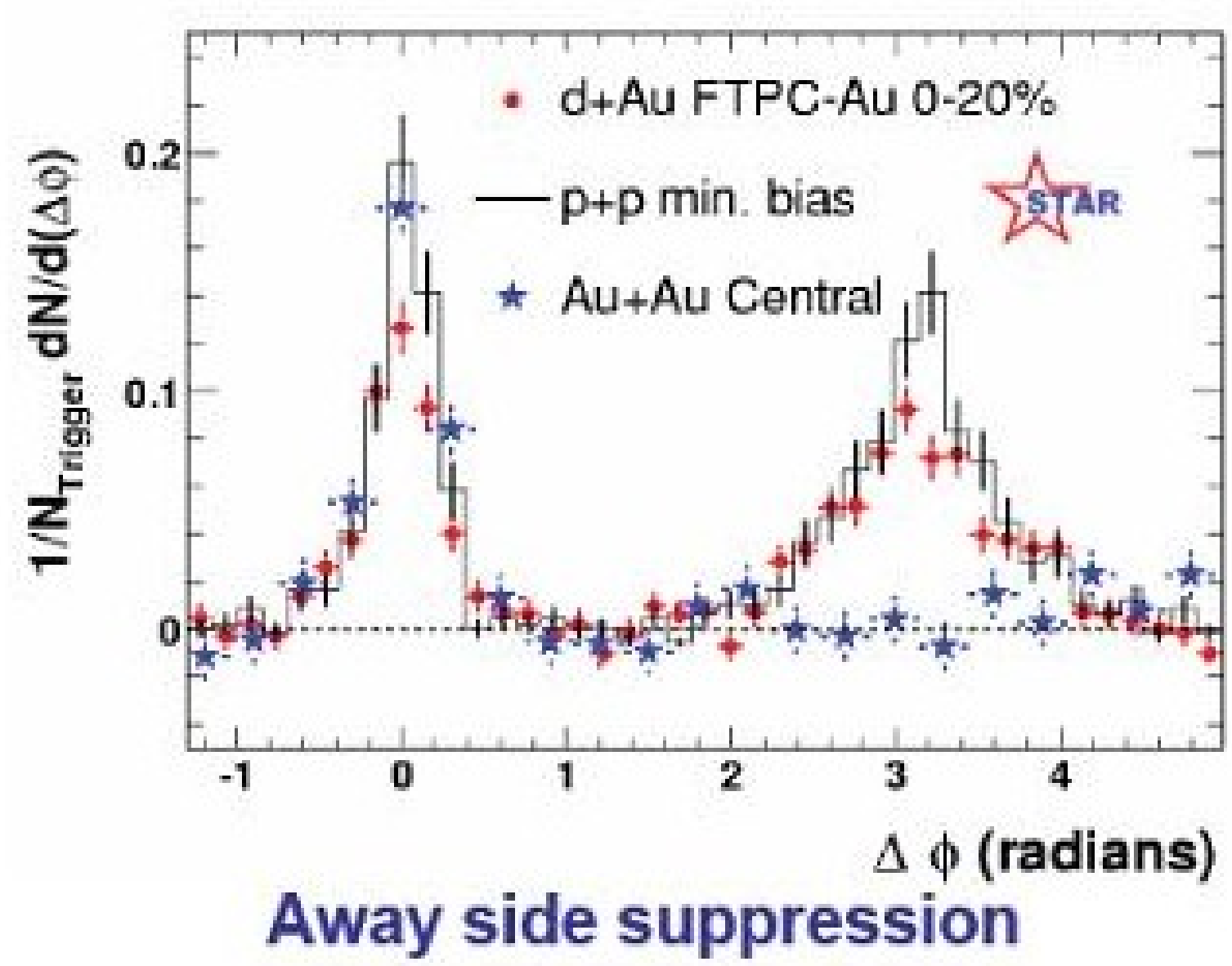

**Fig. 8:** The azimuthal correlations of charged particles relative to a high $p_T$ trigger particle [11]. The jet outgoing in opposite direction for central Au–Au collisions (blue stars) is suppressed compared to the proton–proton (black histogram) and d–Au (red dots).

## 7 The Large Hadron Collider

The Large Hadron Collider (LHC) [13] accelerates protons in a 27 km long tunnel located at the European Organization for Nuclear Research (CERN) in Geneva, Switzerland. The LHC will also accelerate lead ions to make them collide at the highest energy ever.

The acceleration process starts in Linac 2 for protons and Linac 3 for lead ions. The protons accelerated in Linac 2 are injected into a Proton Synchrotron Booster with an energy of 50 MeV. In the synchrotron, protons reach an energy of 1.4 GeV. The Super Proton Synchrotron (SPS) has been modified to deliver a high-brightness proton beam required by the LHC. The SPS takes 26 GeV protons from the Proton Synchrotron (PS) and brings them to 450 GeV before extraction.

The Linac 3 produces 4.2 MeV/u lead ions. Linac 3 was commissioned in 1994 by an international collaboration and upgraded in 2007 for the LHC. The Low Energy Ion Ring (LEIR) is used as a storage and cooler unit providing ions to the (PS) with an energy of 72 MeV/nucleon. Ions will be further accelerated by the PS and the SPS before they are injected into the LHC where they reach an energy of 2.76 TeV/nucleon.

The LHC consists of 1232 superconducting dipole magnets with double aperture that operate at up to 9 Tesla magnetic field. The accelerator also includes more than 500 quadrupole magnets and more than 4000 corrector magnets of many types.

Ions are obtained from purified lead that is heated to 550 $^{\circ}$C. The lead vapour is then ionized with an electric current that produces various charge states. The $Pb^{27+}$ ions are then selected with magnetic fields. This process takes place in an Electron Cyclotron Resonance (ECR) source (Fig. 9).

The ECR lead source is equipped with an hexapole permanent magnet. The plasma chamber is immersed in a solenoidal magnetic field. Pulsed beam currents produce $Pb^{27+}$ ions that are then extracted to the Linac.

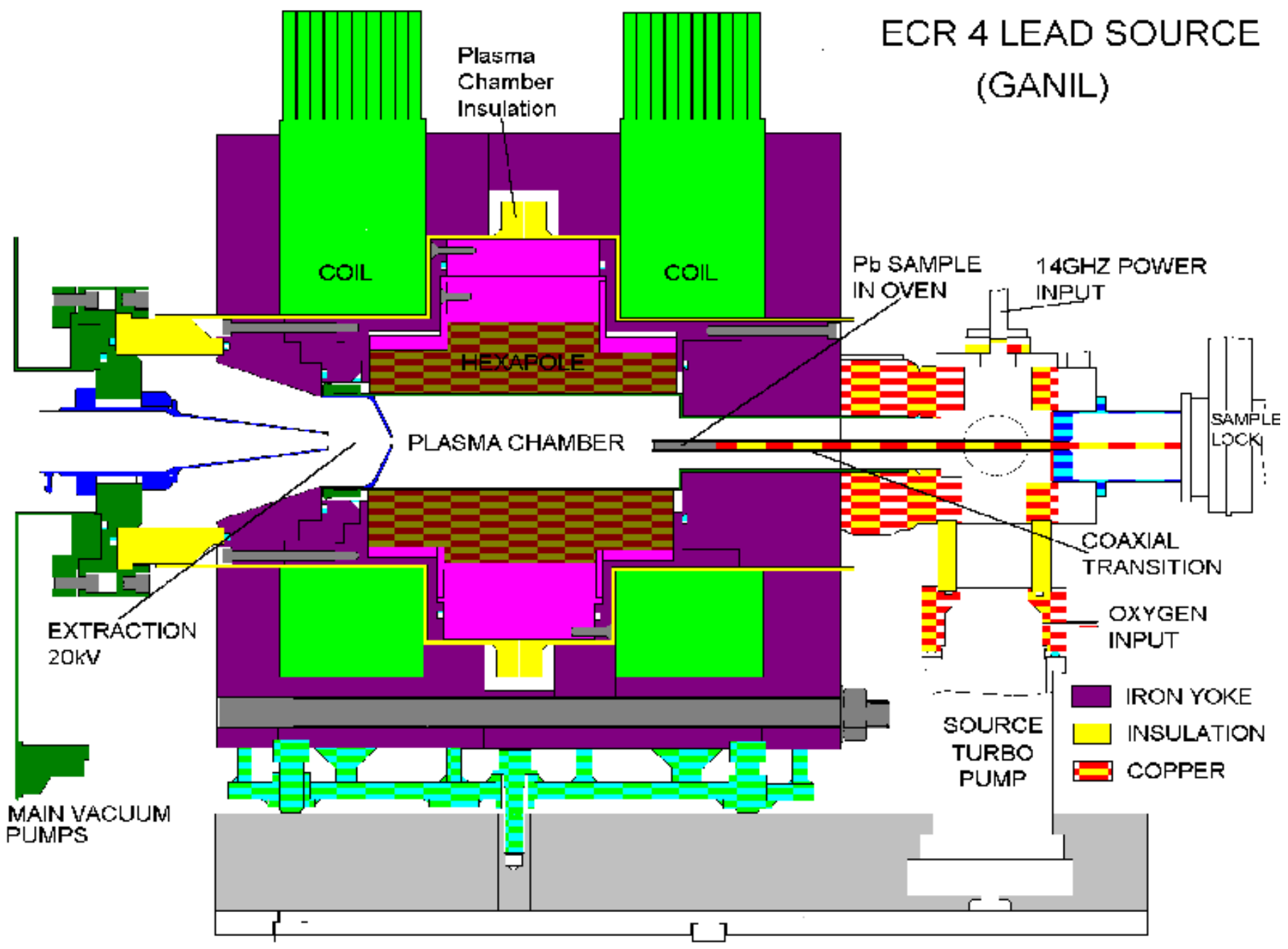

**Fig. 9:** Electron Cyclotron Resonance (ECR) ion source.

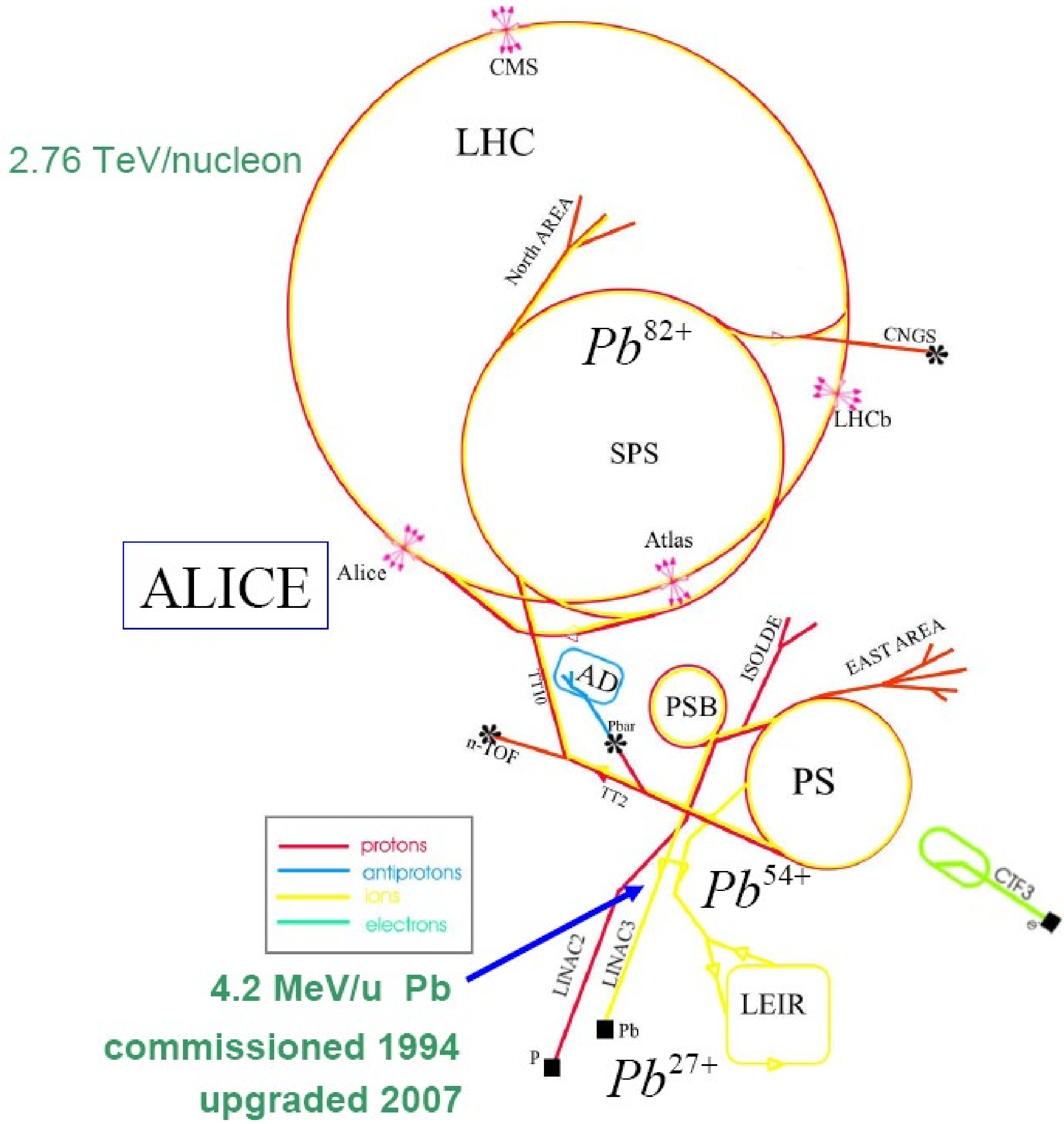

**Fig. 10:** Accelerators at CERN. The process of acceleration starts in Linac 2 and Linac 3 for protons and ions respectively.

After acceleration, the lead ions go through a carbon foil that strips them to $Pb^{54+}$, which are accumulated in the Low Energy Ion Ring (LEIR). LEIR is a circular machine which transforms the long pulses of Linac 3 into high-density bunches needed by the LHC. LEIR injects bunches of ions to the PS.

At the SPS, ions go once more through a thin aluminium foil which strips them to $Pb^{82+}$. The thickness of the stripper foil has to be chosen carefully to reduce contamination of lower charge states and keep emittance low. Foils of 0.5 to 1 mm thickness have been studied. In this way, fully stripped lead ions are obtained for the LHC.

Figure 10 shows the accelerators at CERN that are in use for the LHC.

The total cross-section of proton–proton interaction at 7 TeV could be inferred from hadronic cross-section measurements at lower energy [14]. It would be around 110 mbarn and correspond to about 60 mbarn of inelastic-scattering cross-section. The accelerator, at its design level, will reach a luminosity of $10^{34}\ \mathrm{s}^{-1}\ \mathrm{cm}^{-1}$ which means that the interaction rate will be

$$rate = 10^{34}\,(1/\mathrm{cm}^2\,\mathrm{s}) \times 60 \times 10^{-3}\,\mathrm{barn} \times 10^{-24}\ \mathrm{cm}^2/\mathrm{barn} = 600 \times 10^{6}\ \mathrm{collisions/s}\,.$$

A 25 ns interval between bunches gives a 40 MHz crossing rate. On average 19 inelastic events will occur each time bunches cross. Since there will be gaps in the beam structure, an average crossing rate of 31.6 MHz will be reached. Detectors at the LHC must be designed to cope with these frequencies. However, ALICE will run at a modest 300 kHz interaction rate in proton–proton mode and 10 kHz in Pb–Pb.

During autumn 2009, bunches of protons will be injected into the LHC ring. During the start-up phase, first collisions with protons at 3.5 TeV will take place. An increase of the proton beam energy in a second phase is foreseen. By the end of the run with protons in year 2010, lead-ion collisions will be produced.

The ALICE experiment is ready to take data on all the phases of the accelerator operation.

## 8 A Large Ion Collider Experiment

The ALICE experiment has been designed to observe the transition of ordinary matter into a plasma of quarks and gluons [15]. At the energies achieved by the LHC, the density, the size, and the lifetime of the excited quark matter will be high enough to allow a careful investigation of the properties of this new state of matter. The temperature will exceed by far the critical value predicted for the transition to take place.

ALICE has been optimized to study global event features. The number of colliding nucleons will provide information on the energy density achieved. The measurement of elliptic flow patterns will provide information about thermalization on the partonic level and the equation of state of the system in the high-density phase. Particle ratios in the final state are connected to chemical equilibration and provide a landmark on the trajectory of the system in the phase diagram. The spacetime evolution of the system can be investigated via particle interferometry, complemented by the study of resonaces. Moreover, important information about the system properties can be obtained by the study of hard probes, which will be produced abundantly at the LHC. Deconfinement may be reflected in the abundancies of *J/ψ* and Upsilon. The study of jet production on an event-by-event basis will allow one to investigate the transport properties of hard-scattered partons in the medium, which are expected to be strongly modified if a quark–gluon plasma is formed.

ALICE is also well suited for studies of proton–proton and photon–photon reactions. Photon–photon reactions include QED and QCD processes that go from lepton-pair to hadron and jet production. As for proton–proton interactions, diffractive physics would be an exciting area of research.

The ALICE detector will have a tracking system over a wide range of transverse momentum which goes from 100 MeV/*c* to 100 GeV/*c* as well as particle identification able to separate pions, kaons, protons, muons, electrons, and photons.

A longitudinal view of the ALICE detector is shown in Fig. 11. A detailed description of the ALICE detector can be found in Ref. [16].

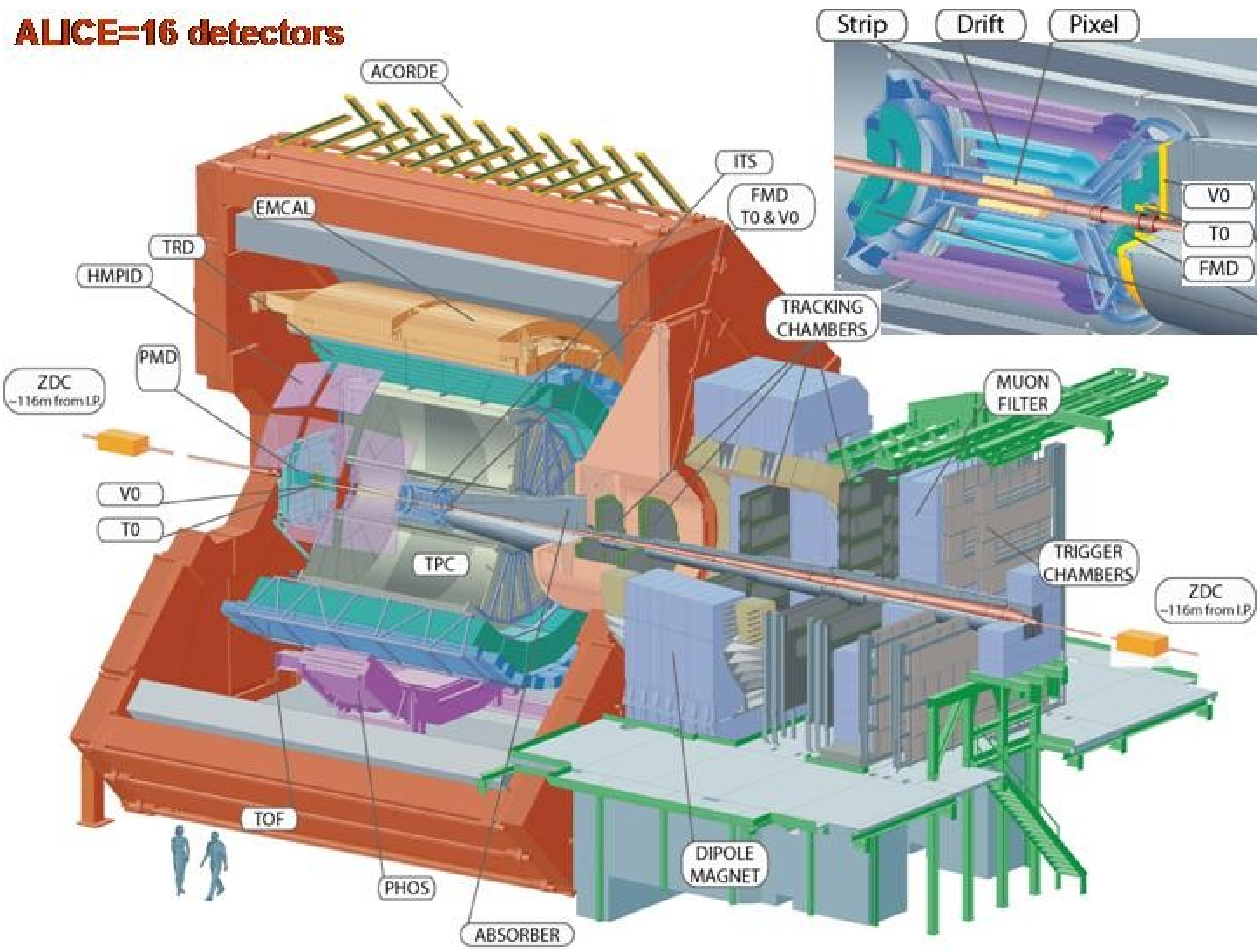

**Fig. 11:** The ALICE experiment consists of 16 detector subsystems. It combines particle identification, tracking, calorimetry, and trigger detectors.

In the forward direction a set of tracking chambers inside a dipole magnet will measure muons. An absorber will stop all the products of the interaction except for the muons which travel across and reach the tracking chambers that form the muon arm.

The central part of the ALICE detector is located inside a solenoid that provides a magnetic field of 0.5 T. The central tracking and particle identification system cover $-0.9 < \eta < 0.9$.

Electrons and photons are measured in the central region: photons will be measured in PHOS, a high-resolution calorimeter 5 m below the interaction point. The PHOS is built from $PBWO_4$ crystals which have a high light output.

The track measurement is performed with a set of six barrels of silicon detectors and a large Time Projection Chamber (TPC). The TPC has an effective volume of 88 $m^3$. It is the largest TPC ever built. These detectors will make available information on the energy loss allowing particle identification too. In addition to this, a Transition Radiation Detector (TRD) and a Time-of-Flight system will provide excellent particle separation at intermediate momentum. The Time-of-Flight system (TOF) uses Multi-gap Resistive Plate Chambers (MRPCs) with a total of 160 000 readout channels. A Ring Imaging Cherenkov detector will extend the particle identification capability to higher momentum particles. It covers 15% of the acceptance in the central area and will separate pions from kaons with momenta up to 3 GeV/*c* and kaons from protons with momenta up to 5 GeV/*c*.

A Forward Multiplicity Detector (FMD) consisting of silicon strip detectors and a Zero Degree Calorimeter (ZDC) will cover the very forward region providing information on the charge multiplicity and energy flow. A honeycomb proportional counter for photon multiplicity (PMD) measurements is located in the forward direction on one side of the ALICE detector.

The trigger system is complemented by a high level trigger (HLT) system which makes use of a computer farm to select events after read-out. In addition, the HLT system provides a data quality monitoring.

The V0 system is formed by two scintillation counters on each side of the interaction point. The system will be used as the main interaction trigger. In the top of the magnet, A Cosmic Ray Detector (ACORDE) will signal the arrival of cosmic muons. We briefly describe these two systems as examples of devices now in operation in the ALICE detector.

## 8.1 The V0 detector

The V0 system consists of two detectors: V0A and V0C, located in the central part of ALICE. The V0A is installed at a distance of 328 cm from the interaction point as shown in Fig. 12, mounted in two rigid half-boxes around the beam pipe. Each detector is an array of 32 cells of plastic scintillator, distributed in 4 rings forming a disc with 8 sectors. For the V0C, the cells of rings 3 and 4 are divided into two identical pieces that will be read with a single photomultiplier. This is done to achieve uniformity of detection and a small time fluctuation.

In proton–proton mode the mean number of charged particles within 0.5 units of rapidity is about 3. Each ring covers approximately 0.5 units of rapidity. The particles coming from the main vertex will interact with other components of the detector generating secondary particles. In general, each cell of the V0 detector will, on average, register one hit. For this reason the detector should have a very high efficiency. In Pb–Pb collisions the number of particles in a similar pseudo-rapidity range could be up to 4000 once secondary particles are included. Comparing the number of hits in the detector for proton–proton versus Pb–Pb mode, we can see that the required dynamic range will be 1–500 minimum-ionizing particles.

The Hamamatsu photomultiplier tubes (PMT) are installed inside the magnet not far from the detector. In order to tolerate the magnetic field, fine mesh tubes have been chosen. The segments of the V0A detector were constructed with a megatile technique [17]. This technique consists of machining the plastic scintillator and filling the grooves with $TiO_2$ loaded epoxy in order to separate one sector from the other.

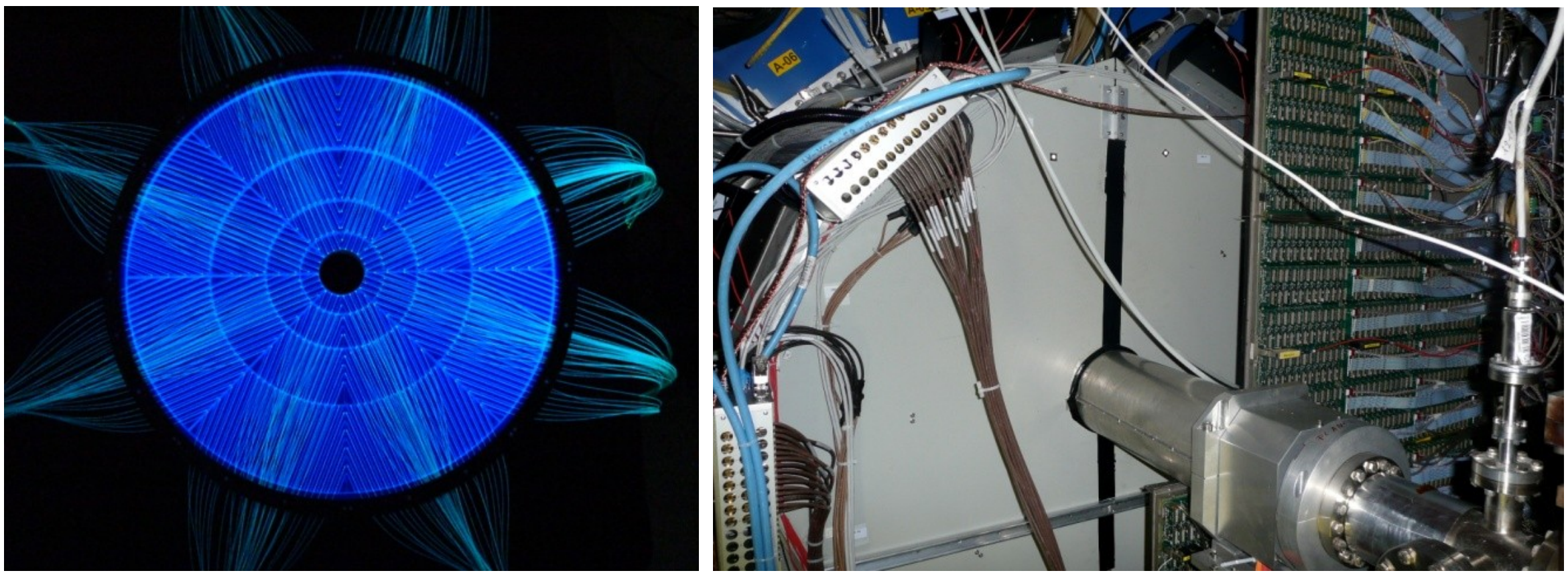

**Fig.12:** The V0A before optical isolation (left). The segmentation and the optical fibres are visible. On the right side, V0A in its box placed in the final position around the beam pipe. One half of the PMD just in front of the V0A can be seen in this picture.

A detailed description of the V0 system can be found in Ref. [18]. Figure 12 shows the V0A detector in its mechanical structure.

### 8.2 A Cosmic-Ray Detector ACORDE

The cosmic-ray detector consists of an array of 60 scintillator counters located in the upper part of the ALICE magnet [19]. The plastic used for the construction of the detector was part of the DELPHI detector. The material was carefully studied and the design of the detector was made according to the capabilities of the plastic available. The material was transported to Mexico where the construction was done.

Each module has a sensitive area of $1.9 \times 0.195\ \mathrm{m}^2$ and is built with two superimposed plastics. The doublet has an efficiency around 90% along the module.

The cosmic-ray detector

- Generates a single muon trigger to calibrate the Time Projection Chamber and other components of ALICE.
- Generates a multi-muon trigger to study cosmic rays with the help of tracking systems like the ITS and the TPC.
- Provides a wake-up signal for the Transition Radiation Detector.

The geometry is shown in Fig. 13. Modules on the far ends of the inner and outer faces of the magnet were moved to the centre of the upper face in order to have a much better efficiency for single muons.

Figure 13 shows a real cosmic-ray event reconstructed with the Time Projection Chamber and projected to ACORDE on the top of the magnet. This event contains 52 muons that fired 38 modules of ACORDE. It was recorded during the cosmic data run in October 2008.

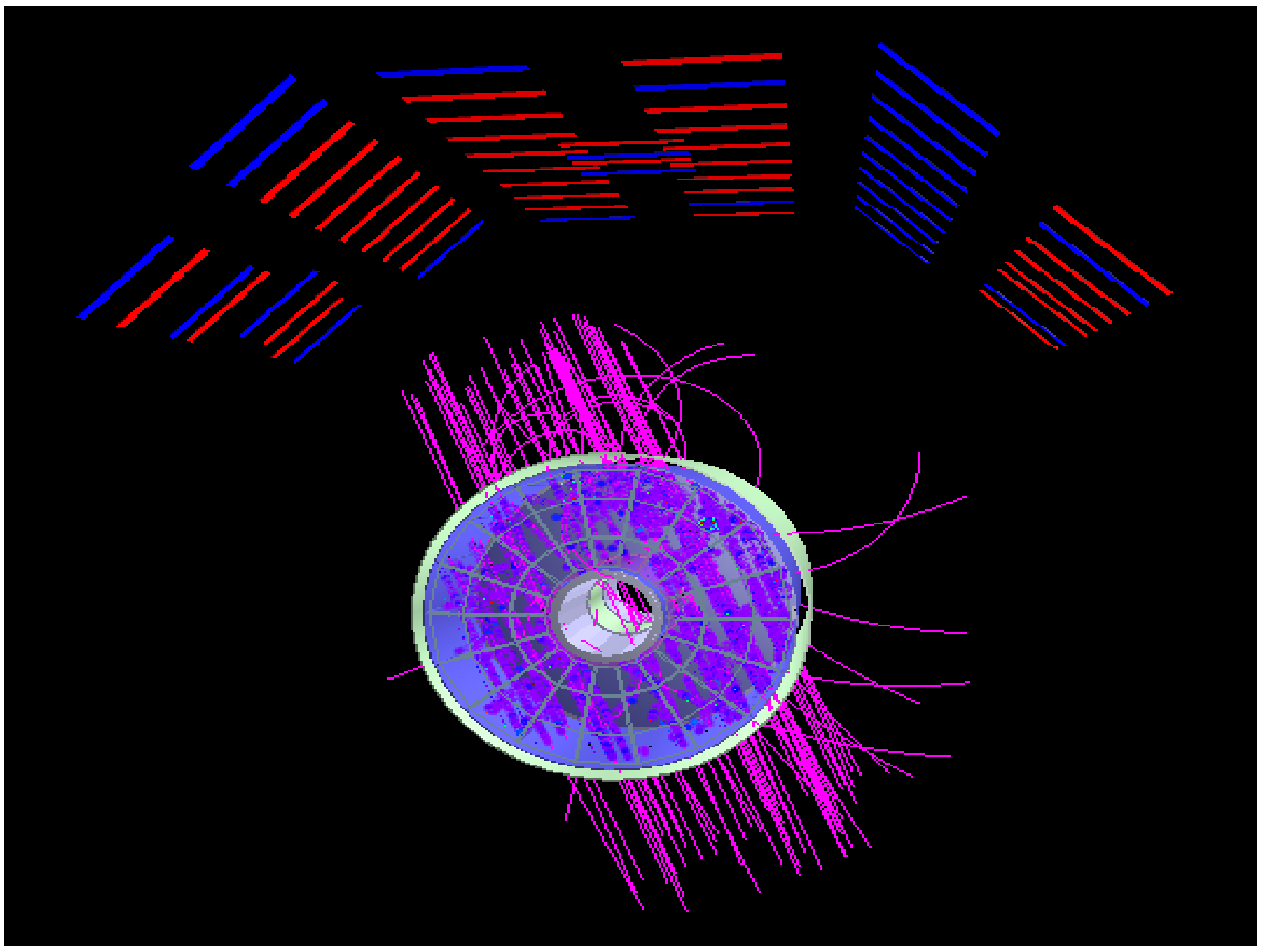

**Fig.13:** ACORDE modules can be seen in the top. This event was taken during a cosmic data run in October 2008. The cosmic-ray detector triggered the TPC to register 52 muons in one single event.

In 2009 a period of two months of cosmic studies will be conducted. The cosmic-ray detector will play a crucial role in triggering interesting events like the one shown here. The commissioning of several systems will be done during this period but interesting physics could be a bonus before accelerator activities start later on this year.

## Acknowledgement

I want to thank the organizers of the School for the kind invitation to deliver these lectures. In particular I would like to thank Danielle Metral-Lillestol and Egil Lillestol for their kind support and warm hospitality.